\documentclass[referee]{raa}
\usepackage{graphicx,times}
\graphicspath{{Fig/}} 
\usepackage{natbib}
\usepackage{amssymb,amsmath}
\usepackage{ulem}

\bibpunct{(}{)}{;}{a}{}{,}

\usepackage[pagebackref=true]{hyperref}

\begin{document}
	
	\title{Magnetic Cloud Boundary Identification Using a Local-Normalized Magnetic Field Parameter}
	 
	\volnopage{ {\bf 20XX} Vol.\ {\bf X} No. {\bf XX}, 000--000}
	\setcounter{page}{1}
	
	\author{{Ziwei Huang}
		\inst{1}, Zhenjun Zhou\inst{1,*}\footnotetext{$*$Corresponding Author.}, Wei Su\inst{2}, Yudong Ye\inst{3}, {Yuming Wang}\inst{3}
	}
	\institute{School of Atmospheric Sciences, Sun Yat-sen University, Zhuhai 519082, People's Republic of China;
		{\it zhouzhj7@mail.sysu.edu.cn}\\
		\and
		TianQin Research Center for Gravitational Physics \& School of Physics and Astronomy, Sun Yat-sen University, Zhuhai 519082, People's Republic of China\\
		\and
		CAS Center for Excellence in Comparative Planetology/CAS Key Laboratory of Geospace Environment, University of Science and Technology of China, Hefei 230026, People's Republic of China\\
		\vs \no
		{\small Received 2026 Month Day; accepted 2026 Month Day}
	}
	
	\abstract{
		Due to the lack of quantitative and reproducible criteria for identifying magnetic cloud (MC) boundaries, we propose a parameter that characterizes short-timescale variability in magnetic field strength. The parameter, referred to as the Local-Normalized Magnetic field parameter (LNM), is defined as $\mathrm{LNM}(t)=\log_{10}\left(B(t)/\langle B \rangle_{\mathrm{5m\text{-}med}}(t)\right)$, where $B(t)$ is the total magnetic field strength and $\langle B \rangle_{\mathrm{5m\text{-}med}}(t)$ is its 5-minute running median ending at time $t$.
		This parameter measures the deviation of the magnetic field magnitude from its local background and reveals a clear contrast between the coherent magnetic structure inside MCs and the more variable ambient solar wind. Based on this parameter, we develop a semi-automated method for MC boundary identification, supported by Time Series Scalogram visualization.		
		We further analyze 76 MC events using power spectral density (PSD) and slab fraction diagnostics. The results show that the dissipation-range spectral index inside MCs ($\sim f^{-2.21}$) is systematically smaller than that outside ($\sim f^{-2.59}$ and $f^{-2.89}$), and the slab fraction is reduced, indicating suppressed small-scale variability and enhanced anisotropy. These results support the applicability of the proposed parameter for MC boundary identification.
		\keywords{Sun: magnetic clouds (MCs) --- Sun: Interplanetary coronal mass ejections (ICMEs) --- Sun: coronal mass ejections (CMEs) --- Sun: Solar Wind}
	}
	
	\authorrunning{Z.-W. Huang et al. }            
	\titlerunning{MC Boundary Identification via Local Variability}  
	\maketitle
	
	%
	\section{Introduction}           
	\label{sect:intro}
	Magnetic cloud (MC) was first introduced by \citet{burlaga1981}. The most significant features of MCs are: (1) enhanced magnetic field strength, (2) smooth rotation of magnetic field strength through a large angle, (3) lower plasma beta value, and lower proton temperature. Besides, other plasma characters can be observed in MCs, such as enhancements of the alpha to proton ratio, the bidirectional suprathermal electron, the $\text{O}^{7+}/\text{O}^{6+}$ ratio $>1.0$ (e.g., \citealt{henke1998}), and average iron charge states $> 12$ (e.g., \citealt{lepri2004}). But such plasma characteristics are usually only used to identify MCs, rather than as parameters to confirm the boundaries of MCs.
	
	There has been no consistency among the criteria used to identify the MC boundary so far \citep{burlaga1995}. \citet{wei2003} analyzed 80 typical MCs from the year 1969 to 2001, and proposed a concept of the MC boundary layer (BL). They characterize the MC boundary not as a simple discontinuity, but as a novel non-pressure-balanced structure comprising distinct inner and outer boundaries \citep{wei2006}. The inner boundaries of the BL are just the beginning and the end of the MC body. The discrimination criteria are similar to the features of the MC mentioned above. The outer boundaries are determined by systematic analyses of the magnetic field and plasma signatures.
	
	A statistical study of 60 BL events by \citet{wei2005}, using WIND/TNR data, revealed the frequent presence of diverse plasma wave activities within the boundary layer. Crucially, these wave modes were distinct from those in both the ambient solar wind and the MC, demonstrating the unique and complex wave environment of the BL. Furthermore, the complexity of MC boundaries is underscored by observational evidence of magnetic reconnection (MR) outflows within these layers, as reported by \citet{voros2021}. As noted by \citet{lepping1990}, the identification of MC boundaries is a matter of subjective judgment. There is still a lack of clear and quantitative standards for determining the boundary of MCs, and the complexity of BL further increases the uncertainty of the subjective judgment of MC boundaries.
	
	As one of the most stable structures in interplanetary space, the factors that disturb MCs have attracted widespread attention and research. \citet{raghav2018} observed inward-propagating Alfv\'en waves within an MC. This observation suggests that such waves may interact with the MC structure and influence its evolution. Recent studies have identified the presence of both Alfv\'en Ion Cyclotron (AIC) waves and Kinetic Alfv\'en Waves (KAWs) within MCs \citep{dhamane2023, kumbhar2024}. These observations indicate that kinetic-scale waves likely play a key role in the plasma heating processes inside MCs.
	
	Despite decades of observational and modeling efforts, MC boundary identification still relies heavily on subjective judgment, often involving a combination of magnetic field rotation, plasma signatures, and empirical experience. A key limitation is the lack of a simple, quantitative parameter that can robustly distinguish the coherent magnetic structure of MCs from the surrounding solar wind.
	
	In this context, a diagnostic based on the intrinsic variability of the magnetic field strength provides a promising alternative, as it directly reflects the degree of magnetic coherence without relying on specific structural assumptions. 
	We define this diagnostic parameter, the Local-Normalized Magnetic field parameter (LNM), as
	\begin{equation}
	\mathrm{LNM}(t) = \log_{10}\left( \frac{B(t)}{\langle B \rangle_{\mathrm{5m\text{-}med}}(t)} \right),
	\end{equation}
	where $B(t) = \sqrt{B_R^2 + B_T^2 + B_N^2}$ is the total magnetic field strength, and $\langle B \rangle_{\mathrm{5m\text{-}med}}(t)$ is the 5-minute running median of $B$ ending at time $t$.
	Unlike conventional fluctuation measures such as $\delta B/B$, this parameter does not explicitly isolate perturbations from a mean field. Instead, it characterizes the relative deviation of the instantaneous magnetic field strength from its local background over a finite timescale. In this sense, it captures scale-dependent variations in $|B|$, effectively highlighting the contrast between the more coherent magnetic structure inside MCs and the more variable ambient solar wind. Based on this parameter, we develop a semi-automated method for MC boundary identification. Furthermore, we perform power spectral density (PSD) and slab fraction analyses to examine the turbulence properties associated with this parameter. These analyses provide supporting evidence that the proposed parameter reflects intrinsic differences in magnetic variability and anisotropy between MC interiors and the surrounding solar wind. Section 2 introduces the parameter, the automated boundary identification method, as well as the PSD analysis and slab fraction calculation. Section 3 analyzes and discusses the statistical characteristics of 76 MC events. Section 4 provides a summary of this study.
	
	\section{Data and Method}
	\label{sect:data_method}
	
	\subsection{Data}
	In this study, all MC events were selected from the HELIO4CAST ICMECATv2.2 catalog \citep{mostl2020}, which gathers information on the in situ CME observations by Solar Orbiter, BepiColombo, Parker Solar Probe (PSP), Wind, STEREO-A, STEREO-B, Juno, MAVEN, Venus Express (VEX), MESSENGER, and Ulysses. The catalog is available at \url{https://helioforecast.space/icmecat}.

	For the identification of MC boundaries, we consistently used magnetic field data in the RTN coordinate system with a time resolution of 1 minute. The subsequent PSD and slab fraction analyses require a higher time resolution: to resolve frequency points up to 4 Hz, a native resolution of at least 8 Hz is necessary. Therefore, we used the highest available time resolution for each spacecraft. Regarding the selection of MC events, we aimed to test the method on as many events as possible across a wide range of heliocentric distances. Hence, we chose events observed by Solar Orbiter, PSP, MESSENGER, and Ulysses, which provide broad coverage in heliocentric distance. In contrast, Wind and STEREO spend most of their time near 1 AU, where MC boundaries have already been extensively studied; we therefore did not include them in this analysis. Furthermore, to ensure that the observed variations in the diagnostic parameter are indeed attributable to MC boundaries and to facilitate subsequent analyses, we applied the following selection criteria to the catalog events:
	\begin{enumerate}
	\item Each candidate must exhibit well-defined MC signatures, characterized by an enhanced and smoothly rotating magnetic field vector over an extended interval.
	\item The minimum MC duration must exceed 3 hours, in order to exclude small-scale flux-rope structures embedded in the ambient solar wind.
	\end{enumerate}

	\subsection{Automated Boundary Determining}
	Due to their highly organized internal structure, MCs exhibit markedly weaker short-timescale variability in $|B|$ than the ambient solar wind. 
	Based on this property, we employ the diagnostic parameter $\mathrm{LNM}(t)$ defined in Eq.~(1).
	The 5-minute window is chosen to capture short-timescale variations while filtering out large-scale trends in the magnetic field, providing a balance between sensitivity to local variability and robustness against long-term variations. 
	
	The window length modulates the sensitivity of the LNM parameter to magnetic field variations. A shorter window (e.g., 2 minutes) causes the running median to closely track the original $B(t)$, resulting in small-amplitude $\mathrm{LNM}(t)$ values; this reduces the contrast between the MC interior and the boundary, making the boundary difficult to identify in events where the internal and external fields are not dramatically different. A longer window (e.g., 30 minutes) makes the denominator overly smooth, causing $\mathrm{LNM}(t)$ to exhibit large fluctuations even inside the MC, which amplifies spurious disturbances and hinders accurate boundary detection. For well-structured MC events, window lengths between 2 and 10 minutes yield similar boundary results. Considering both the data resolution (1 minute) and statistical stability, we adopt a 5-minute window as a balanced choice.

	This parameter quantifies the relative deviation of the magnetic field magnitude from its local background over short timescales. It does not explicitly separate fluctuations from a global mean field, but instead provides a scale-dependent measure of magnetic variability.
	
	Inside MCs, where the magnetic field is more coherent, $|B|$ remains close to its local average, leading to smaller parameter values. In contrast, the ambient solar wind exhibits larger deviations, resulting in higher values. This systematic contrast enables the identification of MC boundaries.

	The proposed boundary identification algorithm requires four input parameters: the LNM time series (denoted as $\mathrm{LNM}(t)$ with 1-minute resolution), the standard deviation threshold $\sigma_{\text{th}}$, the check window length $\Delta t_{\text{win}}$, and the initial reference time $t_{\text{start}}$. 

	The default values are defined as follows. The standard deviation threshold is set to $5\sigma_{\text{core}}$, where $\sigma_{\text{core}}$ is the standard deviation of $\mathrm{LNM}(t)$ computed only over the central two-thirds of the MC interval reported in the ICMECAT catalog. The default check window length is $\Delta t_{\text{win}} = 4$ minutes. The default initial reference time $t_{\text{start}}$ is determined by scanning the LNM series with non-overlapping 90-minute windows: the window with the minimum standard deviation of $\mathrm{LNM}(t)$ is selected, and its central time (window start $+45$ minutes) is taken as $t_{\text{start}}$.

	Starting from $t_{\text{start}}$, the left and right boundaries are both initialized to $t_{\text{start}}$ and then expanded outward independently. For the left boundary, at each step the algorithm examines the data segment of length $\Delta t_{\text{win}}$ ending at the current left boundary, i.e., $[t_L - \Delta t_{\text{win}}, t_L]$. For the right boundary, it examines the segment starting at the current right boundary, i.e., $[t_R, t_R + \Delta t_{\text{win}}]$. If the standard deviation of $\mathrm{LNM}(t)$ over the examined segment exceeds $\sigma_{\text{th}}$, the expansion stops and the current time is recorded as the boundary; otherwise, the boundary is advanced outward by 1 minute ($t_L \leftarrow t_L - 1$ min for the left boundary, $t_R \leftarrow t_R + 1$ min for the right boundary) and the evaluation is repeated. The algorithm terminates when both boundaries are identified or the data limits are reached, returning the interval $[t_L, t_R]$ as the estimated MC boundaries.

	The choice of $5\sigma_{\text{core}}$ as the default threshold is empirical. The conventional $3\sigma$ threshold for distinguishing signal from noise was tested on the same dataset. With $3\sigma$, a substantially larger fraction of the resulting intervals were confined to a small internal interval of the MC rather than covering the expected full extent. In contrast, the $5\sigma$ threshold yields intervals that generally agree better with the catalog boundaries and with visual inspection of the magnetic field profiles. However, $5\sigma$ is not universally optimal: the diversity of MC events means that even with $5\sigma$ some cases require manual adjustment of the threshold. As a default parameter, $5\sigma$ performs considerably better than $3\sigma$ on the present dataset. This empirical observation does not indicate a deficiency of the $\mathrm{LNM}(t)$ parameter itself — the systematic difference in $\mathrm{LNM}(t)$ behavior inside and outside MCs is robust. Rather, it suggests that further improvement of the boundary identification procedure is possible, for example by using an adaptive threshold. Future work with larger event samples may allow a more systematic optimization of the threshold selection, including adaptive or event-dependent criteria.

	The proposed method does not strictly require prior knowledge of the cloud interval from a catalog to estimate $\sigma_{\text{core}}$. An alternative is to compute the standard deviation of $\mathrm{LNM}(t)$ over the 90-minute window that exhibits the minimum standard deviation (the same window used to determine the default starting time $t_{\text{start}}$). Because the interior of a MC typically has very low magnetic field variability, the standard deviation over that window serves as a robust proxy for $\sigma_{\text{core}}$. In practice, replacing the catalog-based $\sigma_{\text{core}}$ with this window-based estimate does not appreciably change the final boundary positions, further demonstrating the intrinsic stability of the $\mathrm{LNM}(t)$ parameter.

	In addition to the standard deviation method, there are various methods that can assist us in identifying MC boundaries based on this parameter, such as Time Series Scalogram. \citet{huang2024} introduced the Gaussianity Scalogram, a visualization technique that reveals fractal-like fine structures in solar-wind time series. They used this technique to reveal the minute-scale fractal fine structures in solar wind time series, and this method is also applicable to hour-scale structures like MCs. Using their Time Series Scalogram, we can visualize the fluctuations of the parameter and help us identify the boundaries of the MC. With this method, when $\mathrm{LNM}(t)$ remains low within an interval, a white pyramid-shaped area will form at the last panel, and the base of the pyramid indicates the start and end time of the interval. Therefore, we can evaluate whether the identified boundaries are consistent with the expected magnetic structure of MCs. Whenever the identified MC boundaries showed substantial deviations from the expected ones, we re-evaluated the boundaries by combining the scalogram and supporting plasma data.
	
\begin{figure}[h]
		\centering
		\includegraphics[width=12.0cm, angle=0]{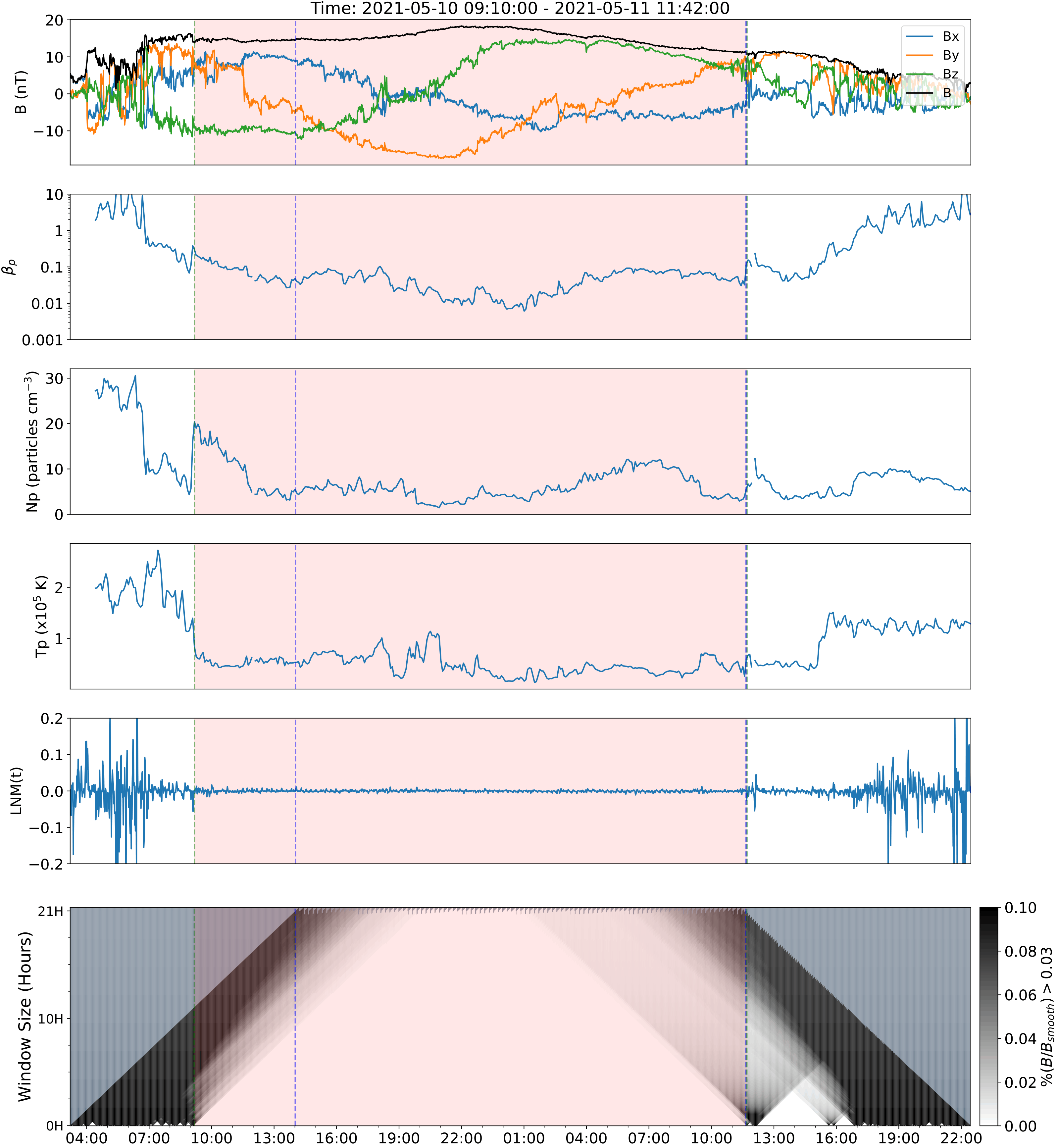}
		\caption{Observations of interplanetary parameters for the MC event on 2021 May 10 observed by Solar Orbiter (MAG instrument, 1-minute resolution). The green dashed lines (left and right) indicate the front and rear boundaries identified by our method, respectively. The blue dashed lines (left and right) indicate the front and rear boundaries from the ICMECAT catalog, respectively. The regions shaded red represent the MC interval determined by our method. The panels from top to bottom represent the total magnetic field strength $B$ and its individual components ($B_x$, $B_y$, $B_z$), the plasma $\beta$, the proton number density ($N_p$), the proton temperature $T_p$, $\mathrm{LNM}(t)$, and the Time Series Scalogram.}
		\label{Fig1}
	\end{figure}

	For example, in the MC event on 2021 May 10 observed by Solar Orbiter, the boundary we obtained differed significantly from the boundary given in the ICMECAT list, so we carefully examined its magnetic field and plasma data, shown in Figure \ref{Fig1}. We can see that the boundary represented by the blue dashed line mainly considers the rotation of the magnetic field, and the x-direction and z-direction components of the magnetic field in front of the MC that do not rotate significantly are not included in the MC part. The boundary determined by $\mathrm{LNM}(t)$ is represented by a green dashed line. Although the boundary is determined solely based on magnetic field data, sudden changes in plasma parameters can be observed at the front boundary. A sudden increase in number density can be observed at the front boundary, which may be due to compression at the front of the MC. The sudden drop in ion temperature is consistent with the definition of the MC itself, and the tail boundary completely coincides with the boundary of the ICMECAT list. 
	Although the interval from approximately 06:00 to 09:00 exhibits relatively low and stable plasma parameters — and the low proton number density is also consistent with typical MC signatures — our method does not identify this interval as part of the MC. Our boundary identification procedure is semi-automated and quantitative. During this interval, the $\mathrm{LNM}(t)$ values fluctuate more vigorously than inside the MC interval, and a clear peak exceeding the preset $5\sigma_{\text{core}}$ threshold occurs exactly at the left green dashed line, which we therefore interpret as the MC boundary. Consistently, the plasma $\beta$ and proton temperature $T_p$ remain relatively high throughout this interval, and clear jumps in these plasma parameters are also observed at the left green dashed line. Thus, excluding this interval from the MC body appears to be justified.
	
	\subsection{Analysis Of Solar Wind And MC Turbulence}
	\subsubsection{Power Spectral Density}
	To determine if the enhanced high-frequency fluctuations at the MC boundary arise from local turbulence, PSD analysis was performed on the magnetic field data. PSD analysis is used to quantify the turbulence properties \citep{youngworth2005, avallone1993}. For solar wind magnetic turbulence, the PSD in the inertial range follows a power-law $P(k) \propto k^{-\alpha}$, where the spectral index $\alpha$ provides a key diagnostic for the turbulent cascade processes. The spectral index ($\alpha$) characterizes the physical processes dominant at different scales. 
	
	The power spectrum of turbulence is conventionally divided into three regions: the energy-containing range at low frequencies, the inertial cascade range at intermediate frequencies, and the dissipation range at high frequencies. Following the framework established by \citet{kiyani2015}, these regions are characterized by distinct spectral index: approximately $-1$ in the energy-containing range ($f < 10^{-4}$ Hz), approximately $-1.65$ in the inertial range ($10^{-3} \text{ Hz} < f < 1 \text{ Hz}$), and approximately $-2.73$ in the sub-ion dissipation range ($1 \text{ Hz} < f < 100 \text{ Hz}$).
	
\begin{figure}[h]
		\centering
		\includegraphics[width=12.0cm, angle=0]{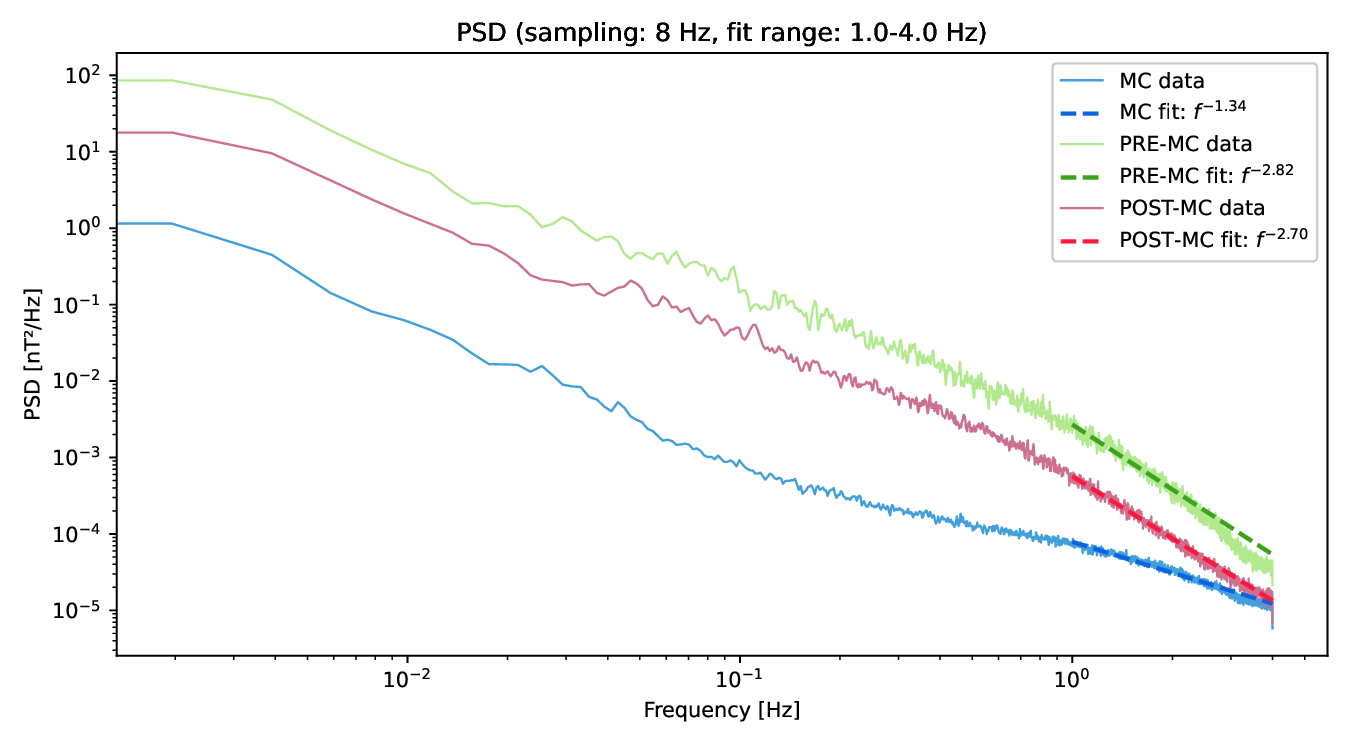}
		\caption{PSD of the MC event on 2021 May 10 observed by Solar Orbiter (MAG instrument, time resolution: 8Hz). The dashed line represents the fitting results of each time period at 1-4Hz.}
		\label{Fig2}
	\end{figure}

	As shown in Figure \ref{Fig2}, the PSD derived from our MC data exhibits a similar feature. We again analyze the MC event of 10 May 2021, observed by Solar Orbiter. Using magnetic field data with a time resolution of 8 Hz, we performed a linear fit to the PSD in the 1-4 Hz frequency band. Our analysis reveals that the spectral index $\alpha$ derived within this MC ($\alpha = 1.34$) is distinctly smaller than that obtained from the pre-MC ($\alpha = 2.82$) and post-MC ($\alpha = 2.70$) intervals. The reduced slope in the dissipation range indicates a relative suppression of small-scale magnetic variability.
	
	\subsubsection{Slab Fraction}
	In solar wind turbulence, the one-dimensional component with wave vectors aligned parallel to the mean magnetic field is termed the "slab" component, a concept initially formulated by \citet{belcher1971}. This component serves as a metric for assessing magnetic field anisotropy within structures like MC. \citet{bieber1994, bieber1996} later advanced a composite "slab/2D" model for solar wind turbulence. This model posits that approximately 85\% of the fluctuation energy in the inertial range resides in the two-dimensional (2D) component, which is characterized by wave vectors predominantly perpendicular to the mean field. Supporting the relevance of this anisotropic framework, \citet{leamon1998} analyzed WIND spacecraft data within CMEs and concluded that the ratio of the 2D to slab component energy increases during MC intervals.
	
	The coordinate system for the wave analysis follows the definition by \citet{bieber1996}. As the magnetic field data are provided in the RTN coordinates (with R being the radial direction from the Sun), we first transform them into a field-aligned coordinate system: the z-axis is defined parallel to the local mean magnetic field direction ($B_0$), the y-axis is given by $-B_0 \times R$, and the x-axis completes the right-handed system as $B_0 \times \hat{y}$. The PSD components along these axes and the slab fraction are subsequently calculated.
	
\begin{figure}[h]
		\centering
		\includegraphics[width=12.0cm, angle=0]{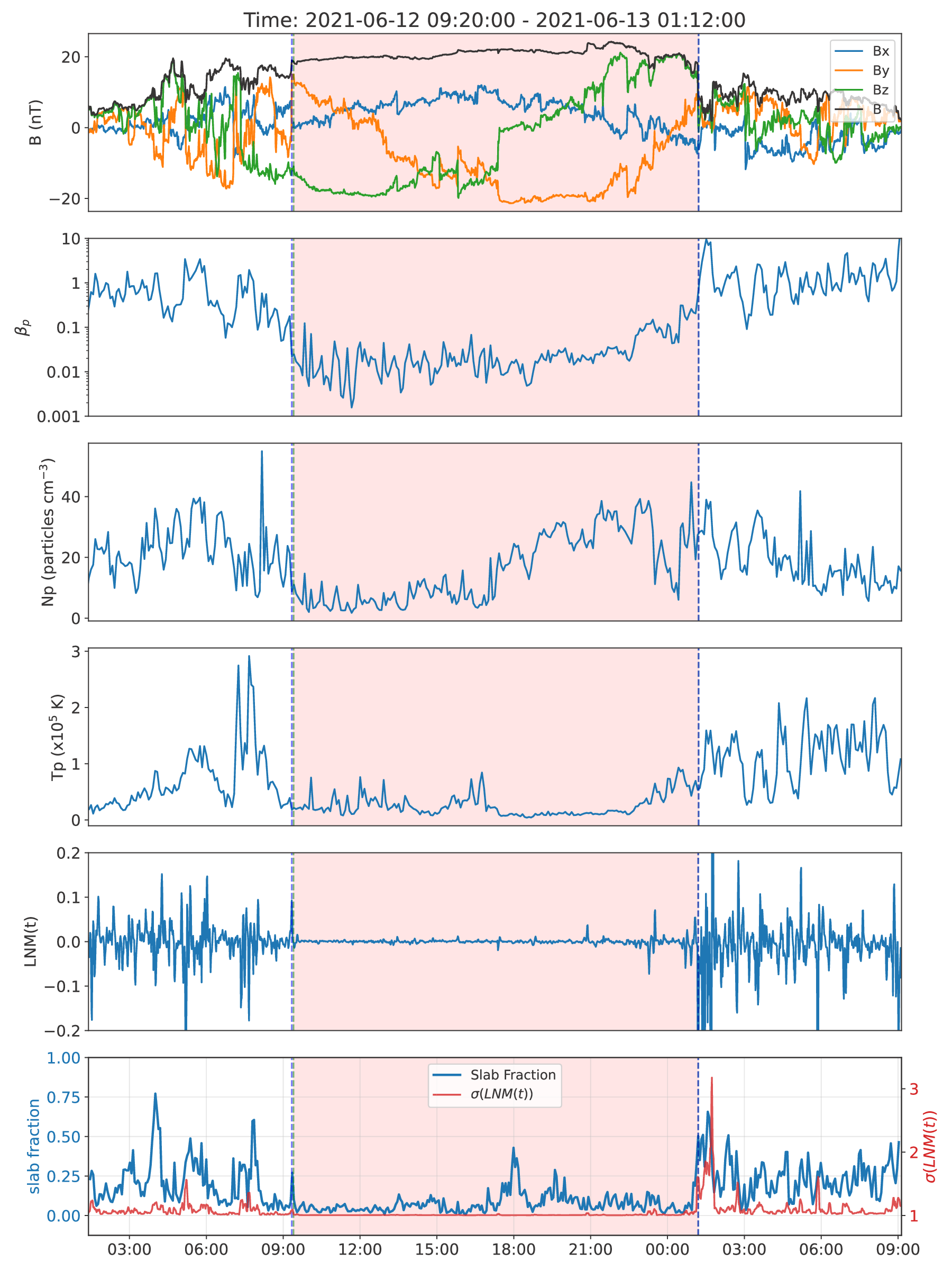}
		\caption{Observations of interplanetary parameters for the MC event on 12 June 2021 observed by PSP (FIELDS magnetometer). The panels from top to bottom represent the total interplanetary field strength $B$ and its individual components ($B_x$, $B_y$, $B_z$), the plasma $\beta$, number density ($N_p$), the proton temperature $T_p$, $\mathrm{LNM}(t)$, and the Slab Fraction with $\sigma_{\mathrm{LNM}(t)}$.}
		\label{Fig3}
	\end{figure}

	To investigate how magnetic field anisotropy within MCs influences boundary identification, we calculated the slab fraction for 76 MC events. Our analysis reveals a notable correlation between the slab fraction and $\mathrm{LNM}(t)$. We illustrate this finding using the MC event observed by the Parker Solar Probe (PSP) on 12 June 2021, with the corresponding magnetic field and plasma parameters displayed in Figure \ref{Fig3}. A pronounced increase in slab fraction is evident near the determined boundaries. In contrast, the slab fraction is generally lower inside the MC. However, we identified a localized region within the MC exhibiting an elevated slab fraction. Examination shows a marked difference in compressibility between the MC's front and tail sections, and this specific region coincides with abrupt changes in both magnetic field and plasma parameters. We speculate that these features may result from the interaction of two distinct ICMEs.
	
\begin{figure}[h]
		\centering
		\includegraphics[width=12.0cm, angle=0]{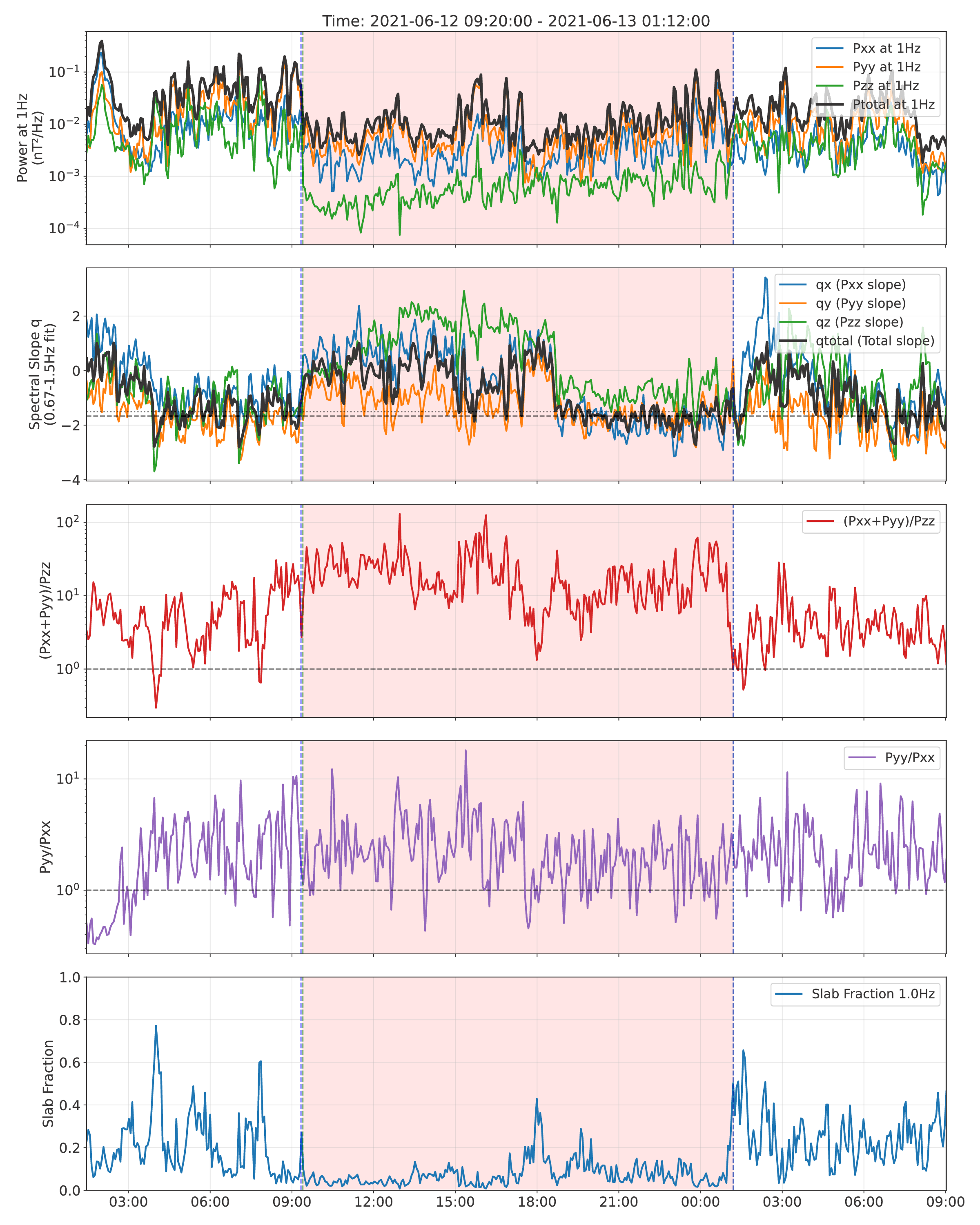}
		\caption{The time profiles of the PSD component ($P$) (top panel), spectral index ($q$) (second panel), ratio of transverse to parallel power ($P_{\perp}/P_{||} = (P_{xx} + P_{yy})/P_{zz}$)(third panel), the $P_{yy}/P_{xx}$ ratio(fourth panel), and the Slab Fraction(bottom panel).}
		\label{Fig4}
	\end{figure}

	The resulting PSD components are presented in Figure \ref{Fig4}. During the MC interval, the PSD for the z-component (parallel to $B_0$) is significantly lower than that for the perpendicular components (x and y). This anisotropy corresponds to the increased value of $(P_{xx} + P_{yy}) / P_{zz}$ shown in the third panel. Furthermore, the second panel reveals two distinct intervals with markedly different spectral slopes. Notably, the slope in the second interval of the MC closely aligns with the Kolmogorov value of $-5/3$.
	
	\citet{leamon1998} observed that in the ambient solar wind, the ratio of perpendicular to parallel power decreases from approximately 10.4 in the inertial range to about 4.9 in the dissipation range, demonstrating a clear frequency dependence of the slab fraction. This necessitates selecting a frequency for our comparative analysis that not only shows a marked difference between the MC interior and the solar wind but also holds clear physical significance. In practice, when computing the slab fraction for all events in our study, we considered a broad range of frequencies from 0.1 Hz up to the Nyquist frequency (half of the sampling rate). For conciseness and clarity, the analysis presented here focuses on the results at 1 Hz.

	\section{Results and Discussion}
	\label{sect:results}
	
	\subsection{Statistical Analysis}
	To determine whether the properties above are common among most MCs, we calculated the PSD and slab fraction for all events in our screened dataset. The screening of MCs was based on the criteria of smooth magnetic field rotation. Notably, this screening process was entirely independent of the parameter $\mathrm{LNM}(t)$, thus eliminating artificial trends introduced by subjective selection.
	
\begin{figure}[h]
		\centering
		\includegraphics[width=12.0cm, angle=0]{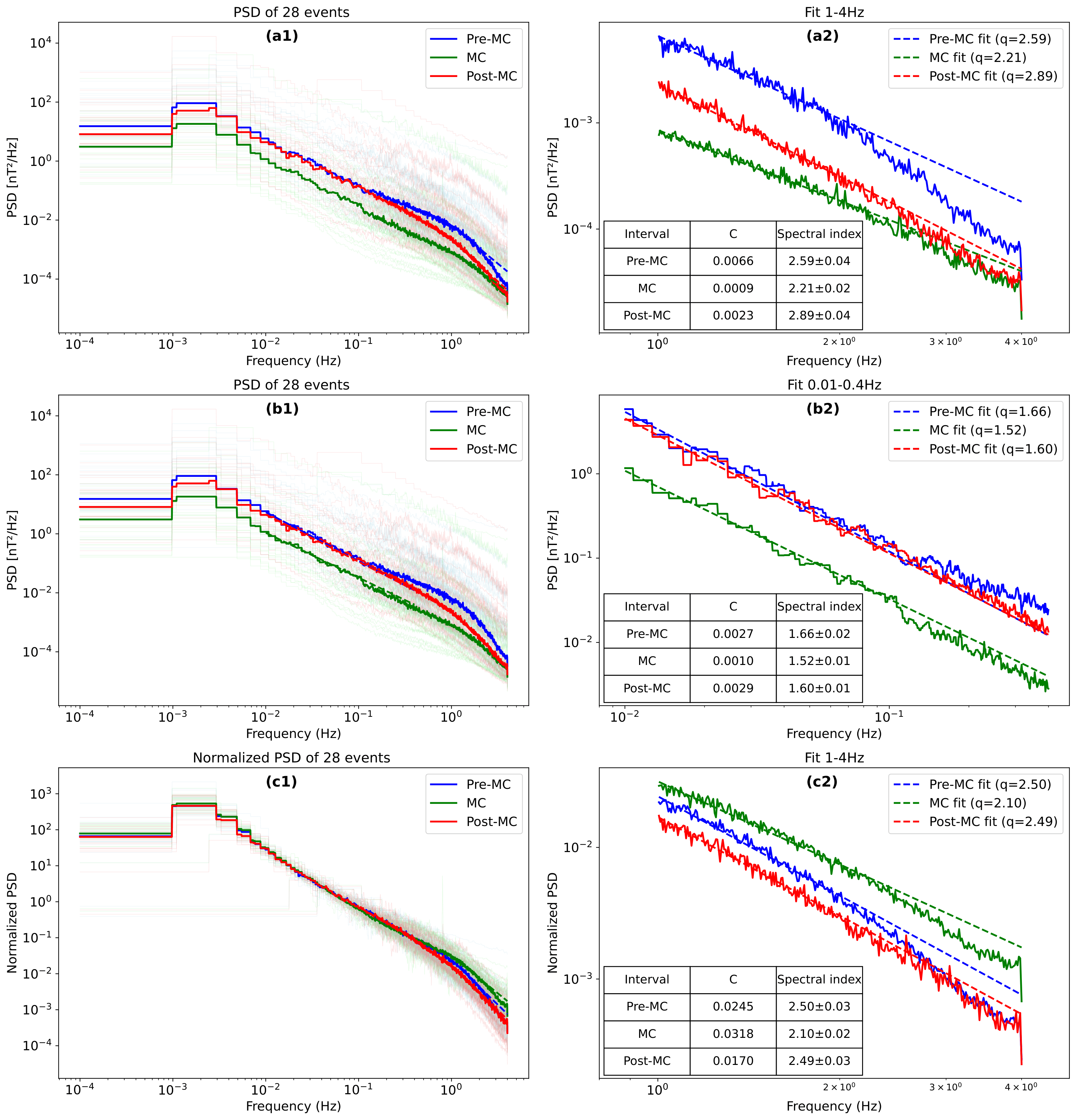}
		\caption{Superposed PSDs of qualifying MC events and their power-law fitting results. (a1, a2) Superposed PSDs of 28 events; the solid line represents the median PSD of the selected events; (b1, b2) Power-law fitting results of the superposed PSDs within the inertial range [0.01-0.4 Hz]; the dashed line denotes the fitting results for each time period. (c1, c2) Superposed PSDs normalized by their median values in the inertial range, along with the corresponding power-law fitting results in the dissipation range [1-4 Hz].}
		\label{Fig5}
	\end{figure}

	For the PSD analysis, which requires fitting the 1-4 Hz dissipation range, we further filtered events to include only those with a time resolution $> 8$ Hz. The PSDs of all qualifying events were then superposed, as shown in Figure \ref{Fig5} (a1) and (a2). Each event was divided into three intervals: Pre-MC (before the MC), MC (within the MC), and Post-MC (after the MC). The left panel shows the median PSD (solid line) against the backdrop of all superposed PSDs. The right panel presents linear fits in 1-4 Hz range, assuming a power-law form $\text{PSD} = C f^{-q}$. The results  reveal that the amplitude factor $C$ is markedly lower during the MC interval compared to the Pre-MC and Post-MC periods. This indicates a more stable magnetic structure and a lower level of Alfv\'enic or other small-scale magnetic fluctuations inside the MC. The concurrently smaller $q$ value suggests a reduced efficiency of the turbulent cascade at small scales within the MC.
	
	In addition, we performed power-law fitting on the inertial range of the superposed PSDs, with the results displayed in Figure \ref{Fig5} (b1) and (b2). The derived spectral index $q$ in the Pre-MC interval is 1.66, which aligns almost perfectly with the classical Kolmogorov scaling of $-5/3$. During the MC interval, the index decreases to $q=1.52$, a value notably close to the prediction of the Iroshnikov-Kraichnan (IK) type turbulence. In the Post-MC interval, the index recovers to $q=1.60$. This intermediate value may characterize the wake region trailing the MC, where residual ordered magnetic fields and velocity shears partially persist, continuing to suppress the turbulent cascade efficiency compared to the undisturbed solar wind.
	
	To isolate the effect of the MC on turbulent scaling from the confounding influence of varying background field strengths, we normalized all PSDs by their median value within the inertial range [0.01-0.4 Hz]. Figure \ref{Fig5} (c1) and (c2) presents the superposed, normalized PSDs and the subsequent power-law fits in the 1-4 Hz range. In the left panel, the normalized PSD for the MC interval is seen to lie above the Pre-MC and Post-MC curves for frequencies above $\sim 0.4$ Hz. Correspondingly, the fitted spectral index $q$ for the MC period is 2.10, which is distinctly smaller than the values of 2.50 and 2.49 obtained for the Pre-MC and Post-MC intervals, respectively. Both observations—the elevated high-frequency power and the flatter spectral slope after normalization—converge to indicate that the turbulent cascade toward smaller scales is less efficient within the MC compared to the ambient solar wind.

	We extended the composite analysis to the slab fraction for 74 MC events (excluding two out of 76 cases with a time resolution less than 2 Hz). To facilitate comparison, the time axis for each event was normalized. The analysis interval for each event was defined from half the MC duration before the front boundary to half the duration after the tail boundary. To facilitate the superposition of all events, we standardized the full analysis interval to the normalized range [0, 1], where the MC interval is situated in the subinterval [0.25, 0.75].
	
\begin{figure}[h]
		\centering
		\includegraphics[width=12.0cm, angle=0]{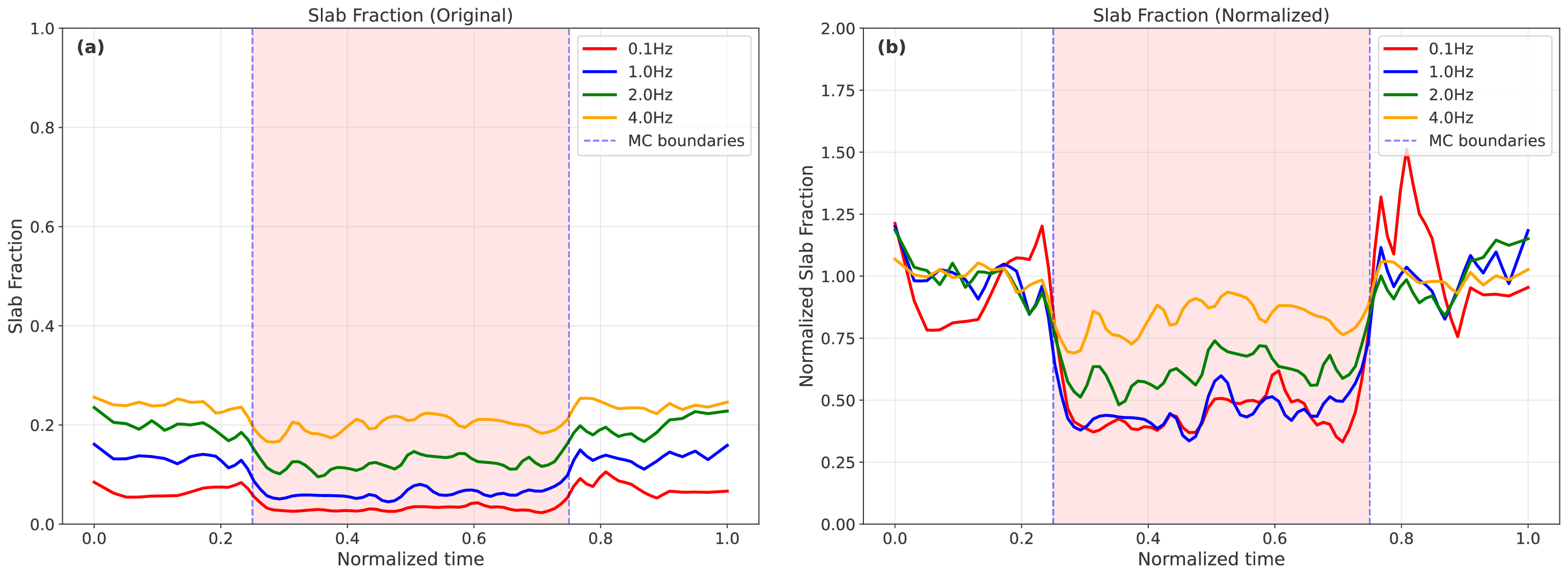}
		\caption{Slab fraction at 0.1-4 Hz versus normalized time: (a) frequency-dependent median slab fraction (colored solid lines) with MC boundaries marked by blue dashed lines; (b) slab fraction normalized to the mean value of the Pre-MC and Post-MC intervals.}
		\label{Fig6}
	\end{figure}

	The slab fractions at different frequencies (0.1 Hz, 1 Hz, 2 Hz, 4 Hz) are shown in Figure \ref{Fig6} (a). The median slab fraction for each frequency is plotted as a solid line in a distinct color, and the MC boundaries are indicated by blue dashed lines. A clear signature emerges across all four frequencies: the slab fraction exhibits a sharp change at the MC boundaries and remains at relatively low values throughout the interior. Furthermore, the overall magnitude of the slab fraction increases with frequency from 0.1 Hz to 4.0 Hz. This trend is consistent with the finding of \citet{leamon1998} that the perpendicular-to-parallel power ratio decreases from the inertial to the dissipation range. Despite the increase in absolute value, the relative change in slab fraction across the cloud boundaries is most pronounced at 1.0 Hz.
	
	To further validate this conclusion, we normalized the slab fraction to the mean value of the Pre-MC and Post-MC intervals (normalized time [0,0.25] and [0.75,1]), as shown in Figure \ref{Fig6} (b). The normalized plots reveal that the relative variation in slab fraction is pronounced at both 0.1 Hz and 1.0 Hz. However, the slab fraction at 0.1 Hz exhibits considerable fluctuations outside the MC boundaries, reducing its contrast and making it less effective for distinguishing between different regions. In contrast, the slab fraction at 1.0 Hz provides a sharper and more stable distinction, with a clearer contrast between the MC interior and the ambient solar wind. This observation reinforces the rationale for selecting the 1.0 Hz slab fraction in our prior individual case analyses.

	\subsection{Discussion}
	
	The parameter proposed in this study provides a quantitative measure of short-timescale variability in magnetic field magnitude relative to a local background. Unlike conventional measures such as $\delta B/B$, it is constructed using a finite-time sliding average and reflects scale-dependent variability rather than fluctuations relative to a global mean field. The reduced values inside MCs indicate a higher degree of magnetic coherence, whereas larger values outside reflect enhanced variability in the ambient solar wind. The choice of the 5-minute timescale does not affect the qualitative distinction between MC and ambient solar wind, as the contrast persists across a range of window sizes. This supports the interpretation that MC boundaries correspond to transitions between regions with distinct variability regimes. Although the statistical consistency across events supports the robustness of the method, the physical mechanisms responsible for the suppression of variability within MCs remain unclear and require further investigation. 
	
	 Given the diverse nature of MCs and their boundary layers, boundary identification has traditionally involved a degree of subjectivity. Weighting factors for magnetic field rotations or plasma parameter jumps often vary between researchers, and visual inspection can still introduce uncertainties on the order of minutes. The method introduced here offers a promising approach to reducing such ambiguities and standardizing the identification process.
	
	Regarding the general applicability of this method, our examination of 76 events revealed that 9 cases did not display a clear pyramidal structure in the Time Series Scalogram. In several of these, the fluctuations exceeded the preset threshold ($B/B_{\text{smooth}} > 0.03$, this threshold follows previous studies and is used here only for comparison purposes), frequently disrupting the expected structure. Importantly, only 3 of the 76 events failed to exhibit a discernible jump in the parameter $\mathrm{LNM}(t)$ near either the front or tail boundary. This indicates that $\mathrm{LNM}(t)$ can serve as a reliable parameter for determining the boundaries of MCs.

	It is worth noting that the 76 events analyzed in this study are predominantly isolated, single-CME cases. Many geoeffective events, however, involve interactions between multiple CMEs, where the boundaries become ambiguous and difficult to resolve. Such interactions can produce complex ejecta with merged magnetic structures, multiple shocks, and highly distorted internal boundaries \citep[e.g.,][]{zhang2007,lugaz2014,liu2024,wang2026}. The present method has not been tested on such interacting events, and its performance under those conditions remains an open question. Since the LNM parameter responds to changes in magnetic-field variability, it may retain sensitivity to large-scale outer boundaries in interacting systems; however, the identification of internal boundaries between multiple CMEs is expected to be substantially more challenging.
	Hence, while the current validation is focused on isolated events, the method's potential for interacting events remains open and warrants future testing.
	
	In practical terms, this criterion has already been applied in our previous work to provide boundary constraints for the Velocity-Modified Uniform-Twist Flux Rope Model with the GH solution \citep{wang2016}, thereby improving the accuracy of the flux-rope axis. Beyond this, the parameter shows potential for use in the automated, real-time detection of MCs and other ICME structures by spacecraft, offering a computationally efficient means for preliminary boundary recognition.
	
	Both statistical analyses and prior observations collectively indicate that fluctuation amplitudes within ICMEs are substantially lower than in the ambient solar wind \citep{zurbuchen2006}. In attempting to elucidate the physical nature of the residual ICME fluctuations, we performed Walén tests and computed the normalized magnetic helicity for our event set. These analyses largely excluded the presence of pure Alfv\'enic or kinetic Alfv\'en wave activity in most events. This finding appears consistent with the scenario proposed by \citet{raghav2018}, wherein the inward propagation and possible dissipation of Alfv\'enic fluctuations may contribute to the evolution of the magnetic structure, ultimately contributing to the MC's erosion and assimilation into the background solar wind. Consequently, the complex evolutionary history of MCs in interplanetary space likely precludes their description by a single, coherent wave mode.
	
	Our work primarily establishes a statistical link—via composite analysis—between short-timescale magnetic variability, quantified by $\mathrm{LNM}(t)$, and key turbulence parameters (PSD and slab fraction). This connection provides a physical basis for the feasibility of our boundary identification method. However, the fundamental physical reason for the suppression of turbulent fluctuations remains an open question, warranting further investigation.

	\section{Summary}
	\label{sect:summary}
	Based on the analysis of 76 MC events, the main contributions of this work are as follows:
	
	\begin{itemize}
		\item  We propose a new method for determining MC boundaries based on short-timescale variability in magnetic field magnitude, complemented by visualization using Time Series Scalograms.
		\item PSD analysis shows that magnetic variability within MCs is reduced compared to the surrounding solar wind. The inertial-range spectral index of 1.52 approached the IK turbulence scaling, while the dissipation-range index of 2.21 was significantly shallower than the post-cloud value of $\sim 2.89$, indicating reduced turbulent cascade efficiency at kinetic scales.
		\item Analysis of the Slab Fraction confirmed a marked anisotropy in the magnetic fluctuations, with power parallel to the mean field being substantially weaker than perpendicular power inside MCs.
		\item The anisotropic signature weakens with increasing frequency. The relative contrast between the MC and ambient solar wind is most pronounced around 1 Hz, which lies near the spectral break between the inertial and dissipation ranges, making it a suitable frequency for detecting this characteristic signature.
	\end{itemize}
	
	\begin{acknowledgements}
	This work is supported by National Natural Science Foundation of China (NSFC) 42274203, Guangdong Basic and Applied Basic Research Foundation (2025A1515010916), and the Specialized Research Fund for State Key Laboratory of Solar Activity and Space Weather. We thank the \textit{PSP, Solar Orbiter, ULYSSES, MESSENGER}  mission teams for providing the spacecraft data used in this study, and Zesen Huang for helpful discussions, and we are grateful to the anonymous referee for constructive feedback that improved the manuscript.
	
	\end{acknowledgements}
	
	\bibliographystyle{raa}
	\bibliography{bibtex} 
	
\end{document}